\documentclass[letterpaper,12pt]{article}

\usepackage[utf8]{inputenc}
\usepackage{fullpage}
\usepackage{amsmath,amssymb}
\usepackage{mathtools}
\usepackage{physics}
\usepackage{url}
\usepackage{todonotes}

\DeclarePairedDelimiterX\innerp[2]{\langle}{\rangle}{#1,#2}

\newcommand{\M}{\vb{M}}

\newcommand{\z}{\vb{z}}

\renewcommand{\r}{\vb{r}}
\newcommand{\R}{\vb{R}}

\newcommand{\beE}{ \begin{equation} }
\newcommand{\enE}{ \end{equation} }
\newcommand{\be}{ \begin{eqnarray*} }
\newcommand{\en}{\end{eqnarray*} }

\newcommand{\ac}{\alpha_{\mathrm c}}

\newcommand{\Rplus}{\R_{+}}
\newcommand{\Rminus}{\R_{-}}

\newcommand{\Reals}{\mathbb{R}}
\newcommand{\eighth}{\frac{1}{8}}

\title{Transition to instability of the leapfrogging vortex quartet}

\author{Roy H. Goodman, Brandon M. Behring}
\date{October 28, 2022}

\begin{document}
\maketitle
\begin{abstract}
The point vortex system is a system of longstanding interest in nonlinear dynamics, describing the motion of a two-dimensional inviscid fluid that is irrotational except at a discrete set of moving point vortices, at which the vorticity diverges. The leapfrogging orbit consists of two rotating pairs of like-signed vortices which, taken as a quartet, propagate at constant velocity. It is known that if the two pairs are initially widely separated, the motion is stable, while if they are closer together it becomes unstable, with this relation represented by a dimensionless  parameter $\alpha$ defined in the text. We here demonstrate analytically that the transition from stability to instability happens at a critical value $\alpha = \phi^{-2}$, where $\phi$ is the golden ratio. This value had been hypothesized based on careful numerics by Tophøj and Aref, and by the present authors using a semi-analytic argument but not previously demonstrated through exact analysis.
\end{abstract}
\section{Introduction}
\label{sec:introduction}

The point-vortex model has a storied history in classical mechanics. Helmholtz derived the system as a simplified model for for a two-dimensional incompressible inviscid fluid in which the vorticity is confined to a discrete set of moving points~\cite{helmholtz}. In this case the equations reduce to a system of ODEs describing the locations of the point vortices evolving over time. 

\begin{figure}[tbp] 
   \centering
   \includegraphics[width=\textwidth]{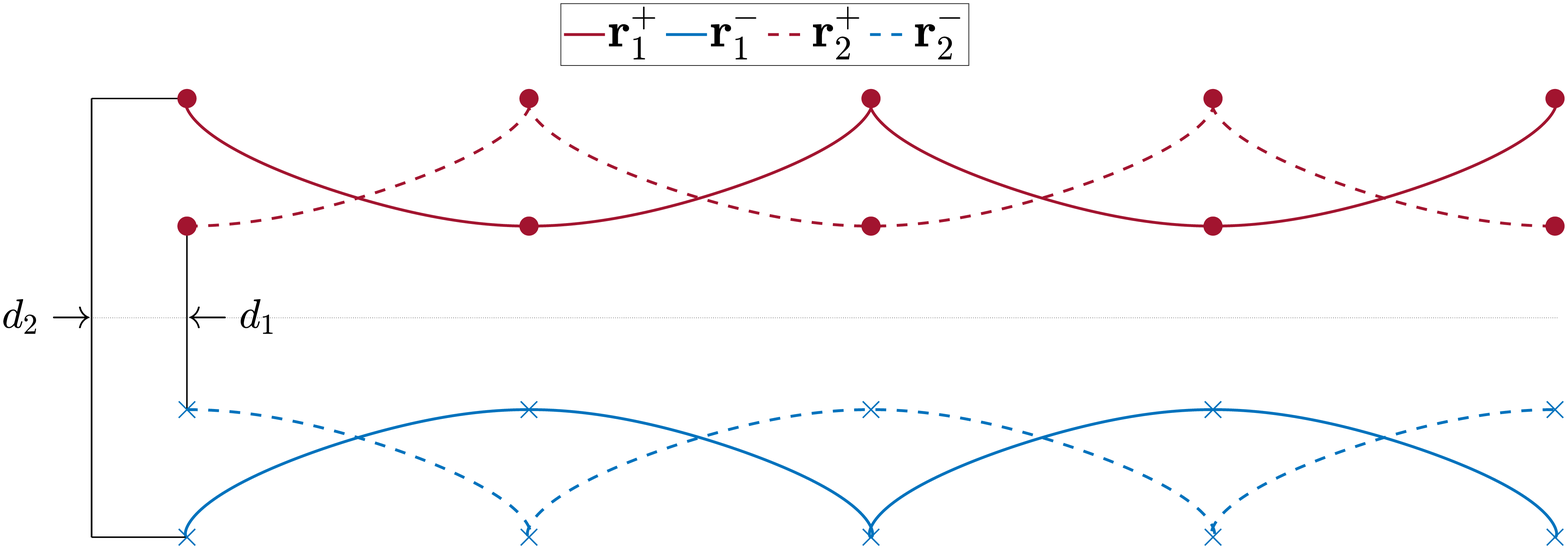}
   \caption{The trajectories of the four vortices in a leapfrog orbit, with initial conditions marked and the distances $d_1$ and $d_2$, as defined in the text, labeled.}
   \label{fig:schematic}
\end{figure}

A system of four vortices of equal strength, two with positive circulation and two with negative circulation, possesses a remarkable family of orbits known as leapfrogging. This was studied, separately, by Love and by Gröbli in the late nineteenth century~\cite{grobli,Love}. The paths of the four vortices in one such orbit are shown in Fig.~\ref{fig:schematic}. Initially, the inner pair travels faster and passes through the outer pair. Subsequently, the inner pair slows and widens, while the distance between the outer pair decreases, causing them to speed up. After half a period of the motion, the identities of the inner and outer pairs are exchanged, and the motion repeats periodically modulo translation. An alternate interpretation is that the motion is hierarchical: the two positive vortices orbit each other, as do the two negative vortices, with the two pairs translating along parallel tracks, while maintaining a mirror symmetry. An analogous motion exists in the motion of a pair of coaxial vortex rings, and the leapfrogging vortex quartet can be seen as a simplified model of this phenomenon.

At the initial time, the four vortices are arranged colinearly, with the two inner vortices separated by a distance $d_1$ and the two outer vortices by a distance $d_2$, with  $\alpha = \frac{d_1}{d_2}$. Both Gröbli and Love determined that such orbits exist for $\alpha_*< \alpha < 1$, where $\alpha_* = 3-2\sqrt{2} \approx 0.171$. For initial conditions with $\alpha\le\alpha_*$ the motion is non-periodic.

\begin{figure}[tbp] 
   \centering
   \includegraphics[width=0.8\textwidth]{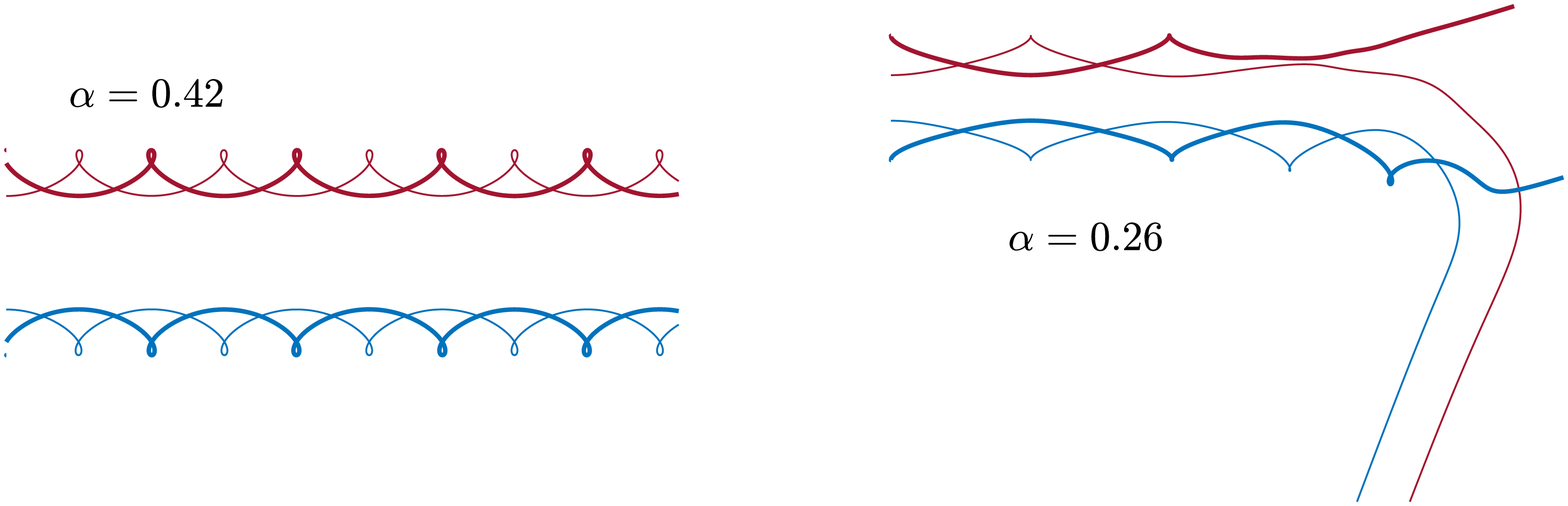} 
   \caption{Two perturbed leapfrogging orbits, the left one stable, the right one stable.}
   \label{fig:2orbits}
\end{figure}

More recently, Acheson noticed, via direct numerical simulation, that the leapfrogging motion is unstable for $\alpha< \ac \approx 0.382, $ and stable for $\alpha>\ac$. A stable motion, with $\alpha\approx0.42$ and an unstable motion, with $\alpha\approx0.26$, are shown in Fig.~\ref{fig:2orbits}. In both simulations, the initial condition is perturbed very slightly from the leapfrogging orbit. The first, which is stable, is not visibly affected by the perturbation, while in the second, which is unstable, the vortices rearrange themselves into a pair of dipoles and escape along oblique trajectories. In the present paper, we confine our discussion to the question of linear stability. However a large variety of \emph{nonlinear} dynamics becomes possible in the unstable regime, with fanciful names like walkabout and braiding orbits, e.g.~\cite{Behring:2020,Tophoj2013,Whitchurch}. We will explore this theme further in an upcoming paper.

Tophøj and Aref made the remarkable observation that $\ac \approx \phi^{-2}$, where $\phi = \frac{1+\sqrt{5}}{2}$ is the golden ratio. They justify this with a formal argument, that the present authors have been unable to follow and believe to be incorrect. That $\ac$ takes such a fortiutous value seems like more than simple coincidence, and the authors' previous paper~\cite{BehringGoodman} documents their initial attempt to prove it. There, we devised a perturbative procedure which allowed us to approximate $\ac$ with increasing accuracy, without ever solving the ODE system numerically. Instead, we used the method of harmonic balance to construct a sequence of matrices, of increasing dimension, each depending on $\alpha$. The determinants of these matrices are polynomials in $\alpha$, and their roots yield an approximation to $\ac$. We constructed these matrices symbolically in Mathematica and found the roots of the polynomials numerically, which confirmed the value of $\ac$ to sixteen digits before we decided to halt it. This did not quite achieve the authors' original goals of proving the specific  value of $\ac$.

In this note, we complete the result. That is, we use mathematical perturbation theory to demonstrate the existence of a bifurcation at $\ac = \phi^{-2}$. To accomplish this result we perform three sequential transformation of different types. First, we apply a sequence of canonical transformations that, taking advantage of conserved quantities, reduces the number of degrees of freedom from four to two. Second, we nonlinearly rescale the Hamiltonian itself in order to desingularize the dynamics in the region of interest. Finally, we change independent variables, which allows us to write down the stability problem in exact form, even though no closed-form solution exists to the original system of equations. We then rely on a somewhat miraculous exact solution to a variable-coefficient linear system, and an application of Floquet theory. 

The remainder of the paper is organized as follows.
In Sec.~\ref{sec:transformations} we set up the equations of motion and review the arguments from our earlier work~\cite{BehringGoodman} in which we transform the problem into a simplified form amenable to analysis. 
Sec.~\ref{sec:linearization} discusses the linearization and a change of independent variables that allows further analysis.
Sec.~\ref{sec:floquet} provides a short review of Floquet theory and and describes a perturbation scheme applicable to the Floquet problem at hand.
In Sec.~\ref{sec:analysis} we finish the analysis that determines the change of stability.

\section{The equations of motion and their transformation}
\label{sec:transformations}
The point vortex model is most easily analyzed by posing it in a Hamiltonian form due to Kirchhoff~\cite{kirchhoff1876}. Consider a system of $N$ vortices with positions $\r_i =(x_i,y_i)$ and circulations $\Gamma_i$. The Hamiltonian is given by
\begin{equation}\label{nvortexham}
    H(\r_1,\ldots,\r_N)=- \sum_{i< j}^N  \Gamma_i \Gamma_j   \log \norm{ \r_i-\r_j }^2,
\end{equation} 
 with equations of motion
\begin{equation}\label{xyode}
    \Gamma_j \dv{x_j}{t} = +\pdv{H }{y_j} \qand 
    \Gamma_j \dv{y_j}{t} = -\pdv{ H }{x_j}, 
    \, j = 1, \ldots, N.
\end{equation}
This Hamiltonian construction is slightly non-standard, as evidenced by the factor of $\Gamma_j$ multiplying the time derivative terms.

We specialize to the case of two vortices of circulation $\Gamma=1$ located at positions $\r_1^+$ and $\r_2^+$, and two of circulation $\Gamma=-1$ located at positions $\r_1^-$ and $\r_2^-$, which has the Hamiltonian
\begin{equation}\label{DimerHam}
   \begin{split}
    H(\r_1^-,\r_1^+,\r_2^-,\r_2^+)
    =&- \log \norm{\r_2^{+}-\r_1^{+} }^2
    - \log \norm{\r_1^{-}-\r_2^{-} }^2\\
    &+ \log \norm{\r_1^{-}-\r_1^{+} }^2
    + \log \norm{\r_2^{-}-\r_1^{+} }^2
     + \log \norm{\r_1^{-}-\r_2^{+} }^2
     + \log \norm{\r_2^{-}-\r_2^{+} }^2.
\end{split}
\end{equation}
We make a symplectic change of variable to coordinates describing the centers of vorticity $\r_\pm$ and for the displacements $\R_\pm$ within the positive and negative pairs
$$
\r_+ = \frac{\r_1^+ + \r_2^+}{2}, \,
\r_- = \frac{\r_1^- + \r_2^-}{2}, \,
\Rplus = \r_1^+ - \r_2^+,  \,
\Rminus = \r_1^- - \r_2^-,
$$
under which the Hamiltonian becomes
\begin{align*}
H(\Rplus,\Rminus,\r_+,\r_-) = & -\log{\norm{\Rplus}^2} 
- \log{\norm{\Rminus}^2} \\
& + \log{\norm{\Rplus - \Rminus +2(\r_+ -\r_-)}^2}
+ \log{\norm{\Rplus - \Rminus -2(\r_+ -\r_-)}^2} \\
& + \log{\norm{\Rplus + \Rminus +2(\r_+ -\r_-)}^2}
+ \log{\norm{\Rplus + \Rminus -2(\r_+ -\r_-)}^2}.
\end{align*}
This depends on $\r_+$ and $\r_-$ only through the combination $\M = 2(\r_+ -\r_-)$. In fact, as show in Ref.~\cite{Tophoj2013}, the components of $\M$ are conserved, and correspond to the vector-valued impulse of the system. 
This yields our final form of the Hamiltonian
\begin{align*}
H(\Rplus,\Rminus) = & -\log{\norm{\Rplus}^2} 
- \log{\norm{\Rminus}^2} \\
& + \log{\norm{\Rplus - \Rminus +\M}^2}
+ \log{\norm{\Rplus - \Rminus -\M}^2} \\
&+ \log{\norm{\Rplus + \Rminus +\M}^2}
+ \log{\norm{\Rplus + \Rminus -\M}^2}.
\end{align*}

Without loss of generality, we can choose $\M=(2,0)$, which amounts to a rotation and scaling of the initial conditions. We then substitute in components, writing
in a standard canonical form \(H(q_+,q_-,p_+,p_-)\) by introducing the components,
\begin{equation}\label{dimer_pq}
    \Rplus = (q_+, p_+) 
    \quad \textrm{and}\quad
    \Rminus = (q_-,-p_-)  .
\end{equation}
The choice of minus sign on $p_-$ normalizes the Poisson brackets so that the evolution equations take the familiar form
$$
\dv{q_j}{t} = \pdv{H}{p_j}, \, \dv{p_j}{t} = -\pdv{H}{q_j}, 
$$
removing the dependence on $\Gamma_j$ seen in system~\eqref{xyode}.

A final change of variables
$$
Q_1=\frac{1}{\sqrt{2}} \left(q_++q_-\right), 
    Q_2=\frac{1}{\sqrt{2}} \left(q_+-q_-\right),
    P_1=\frac{1}{\sqrt{2}} \left(p_++p_-\right),
    P_2=\frac{1}{\sqrt{2}} \left(p_+-p_-\right)
$$
puts this system in the form used by Tophøj and Aref,
\begin{equation}
\begin{split}
    H(Q_1,Q_2,P_1,P_2)=&
    \phantom{+} 
     \log \left((Q_1+Q_2)^2+(P_1+P_2)^2\right)
    + \log \left((Q_1-Q_2)^2+(P_1-P_2)^2\right)\\
    &- \log \left(Q_1^2+(1-P_2)^2\right)
     - \log\left(Q_1^2+(1+P_2)^2\right)\\
    &- \log \left(Q_2^2+(P_1-1)^2\right)
     - \log \left(Q_2^2+(P_1+1)^2\right).
\end{split}
\end{equation}

In these coordinates, the plane $Q_2=P_2=0$ is invariant. This invariant plane includes all the leapfrogging orbits, and this makes the coordinates useful for studying the stability of the leapfrogging orbit. Within this plane, the Hamiltonian simplifies to 
$$
H(Q_1,P_1) =
 2 \log \left(P_1^2+Q_1^2\right)
 -2\log \left(1-P_1^2\right)
 -2 \log \left(Q_1^2+1\right).
$$

As $(Q_1,P_1) \to (0,0)$, the Hamiltonian and its derivatives diverge, corresponding to the divergence in the rotation rate of the two like-signed pairs. This singularity prevents the straightforward application of perturbation theory, but we may desingularize the dynamics by introducing a new Hamiltonian
$$
\tilde{H} = f(H) = \frac{1}{2} e^{H/2}.
$$
Since $\dv{\tilde{H}}{p} = f'(H)\dv{H}{p}$ and $\dv{\tilde{H}}{q} = f'(H)\dv{H}{q}$, and $H$ is constant on trajectories, the orbits of the transformed Hamiltonian system coincide with those of the original system, up to a reparameterization in time.

\begin{figure}[htbp] 
   \centering
   \includegraphics[width=0.75\textwidth]{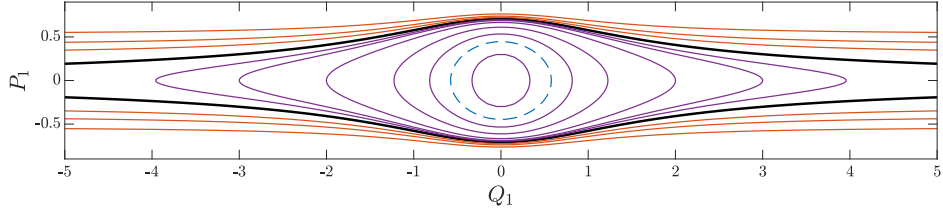} 
   \caption{Level sets of the reduced Hamiltonian function~\eqref{Htilde1dof}, showing periodic (leapfrogging) orbits in purple, unbounded orbits in red, and the critical orbit as a dashed blue line.}
   \label{fig:reducedLeapfrogging}
\end{figure}

The transformed Hamiltonian is
\begin{equation}
\label{Htilde2dof}
\tilde{H}(Q_1,Q_2,P_1,P_2)=
\frac{1}{2} 
\left(\scriptstyle{
 \frac 
  {\left(\left(P_1-P_2\right)^2+\left(Q_1-Q_2\right)^2\right) \left(\left(P_1+P_2\right)^2+ \left(Q_1+Q_2\right)^2\right)}
  {\left(\left(1-P_2\right)^2+Q_1^2\right) \left(\left(1+P_2\right)^2+Q_1^2\right) \left(\left(1-P_1\right)^2+Q_2^2\right) \left(\left(1+P_1\right)^2+Q_2^2\right)}
}
\right)^{\frac{1}{2}},
\end{equation}
and the Hamiltonian on the invariant plane $Q_2=P_2=0$ takes the especially simple form
\begin{equation}
\label{Htilde1dof}
\tilde{H}(Q_1,P_1)= -\frac{1}{2} \frac{1}{1+Q_1^2} + \frac{1}{2} \frac{1}{1-P_1^2}.
\end{equation}
At small amplitude, this has expansion
\begin{equation}\label{harmonic}
\tilde{H}(Q_1,P_1) = \frac{Q_1^2}{2} + \frac{P_1^2}{2} + \ldots,
\end{equation}
so that, to leading order, the motion is simple harmonic with unit frequency. 
Fig.~\ref{fig:reducedLeapfrogging} shows the trajectories due to Hamiltonian~\eqref{Htilde1dof}.  The orbit with ratio $\alpha$ corresponds to the $(Q_1,P_1)$ periodic orbit with $h = \frac{{(1-\alpha)}^2} {8 \alpha}$. The limit $\alpha \to 1^{-}$ corresponds to $h\to 0^{+}$. Periodic orbits exist for $0<h<\frac{1}{2}$ and correspond to leapfrogging orbits in the original system. Most importantly, the value $\ac = \phi^{-2}$ corresponds to $h = \eighth$, and our task is now to show that periodic orbits are linearly stable for $0<h<\eighth$ and linearly unstable for $h>\eighth$.

As suggested by expansion~\eqref{harmonic}, we  will find it useful to make the change of variables 
\begin{equation} \label{polars}
Q_1 = \sqrt{2\rho} \sin{\theta}, \,  P_1 = \sqrt{2\rho} \cos{\theta},
\end{equation}
to the action-angle variables of the simple harmonic oscillator, which puts the Hamiltonian~\eqref{Htilde1dof} in the form
\begin{equation}\label{1DoFHamiltonianPolar}
\tilde{H}(\rho,\theta)=\frac{2 \rho}{2-\rho^2-4 \rho \cos{2 \theta} +\rho^2 \cos{4 \theta} }.
\end{equation} 
Since this transformation is canonical, the new coordinates satisfy
\begin{equation}
\label{thetadot}
\dv{\theta}{t} = \pdv{\tilde{H}}{\rho}
\qand
\dv{J}{t} = - \pdv{\tilde{H}}{\theta}.
\end{equation}

\section{The linearization and a change of independent variable}
\label{sec:linearization}
The next step is to linearize the Hamiltonian system~\eqref{Htilde2dof} about periodic orbits of system~\eqref{Htilde1dof}. The difficulty is that, while this system is formally integrable by quadratures, integration yields a complicated formula for $t(Q_1,h)$ that contains both elliptic integrals and algebraic functions, given in~\cite{BehringGoodman}. This formula can not be analytically inverted, so we seek an alternative method. The cited reference also contains a formula for the period of these motions.

Supposing that the periodic orbits of system~\eqref{Htilde1dof} were known, we linearize system~\eqref{Htilde2dof} by substituting
\begin{align*}
Q_1 & = \sqrt{2\rho} \sin{\theta} + \epsilon x,  & Q_2 & = \epsilon u, \\
P_1 & = \sqrt{2\rho} \cos{\theta} + \epsilon y, & P_2 & = \epsilon v.
\end{align*}
into the evolution equations defined by Hamiltonian~\eqref{Htilde2dof} and keeping the terms linear in $\epsilon$. The $(x,y)$ motion decouples from the $(u,v)$ motion yielding a pair of linear problems, each with two dependent variables. The former is generically neutrally stable, as perturbation within the invariant plane merely leads to an initial condition on a nearby periodic orbit, and, thus, linear separation in time. The interesting linearized motion in the $(u,v)$ coordinates is
\begin{equation}
\label{uvlinearized}
\dv{t} \begin{pmatrix} u \\ v \end{pmatrix} = A(\theta,\rho) \begin{pmatrix} u \\ v \end{pmatrix},
\end{equation}
where
\begin{equation}
A(\theta,\rho) = 
\begin{pmatrix}
 -\frac{\sin {2\theta}}{\rho^2 \cos^2{2\theta}-2 \rho  \cos {2\theta}-\rho^2+1} & \frac{\rho  \left(-\cos^2{2\theta}+3 \cos {2\theta}-2\right)+\cos {2\theta}}{(\rho  \cos {2\theta}-\rho -1)^3} \\
 \frac{\rho  \left(-\cos^2{2\theta}-3 \cos {2\theta}-2\right)+\cos {2\theta}}{(\rho  \cos {2\theta}+\rho -1)^3} & \frac{\sin {2\theta}}{\rho^2 \cos^2{2\theta}-2 \rho  \cos {2\theta}-\rho^2+1} \\
\end{pmatrix}
 \label{Adef}
\end{equation}

While $(\theta(t),\rho(t))$ are not obtainable in closed form, we can rewrite system~\eqref{uvlinearized} explicitly by changing the independent variable from $t$ to $\theta$, since $\theta$ increases monotonically on trajectories. This idea goes back to Newton's proof that the bodies in a two-body gravitational system trace elliptical orbits~\cite{Jose}. We first solve equation~\eqref{1DoFHamiltonianPolar} for $\rho$ in terms of $\theta$ and the energy level $\tilde{H}=h$, finding
\begin{equation}
  \rho = 
   \frac{1+2 h \cos{2 \theta} - \sqrt{1+4 h^2+4 h \cos{2 \theta} }}{h (-1+\cos{4 \theta})}.
\end{equation}
The apparent singularity in this expression at the vanishing of the denominator is removable, as the numerator vanishes to the same order. There are many such apparent singularities in the calculation that follows, all of them removable.

We rewrite the linear system~\eqref{uvlinearized} with $\theta$ as the independent variable  using the chain rule and Eq.~\eqref{thetadot},
$
\dv{t} = \dv{\theta}{t} \dv{\theta} = \pdv{\tilde{H}}{\rho} \dv{\theta},
$ 
yielding
\begin{equation}
\label{uvlinearizedtheta}
\dv{\theta} \begin{pmatrix} u \\ v \end{pmatrix} = 
\left(\pdv{\tilde{H}}{\rho}\right)^{-1}A\left(\theta,\rho(\theta,h)\right) 
\begin{pmatrix} u \\ v \end{pmatrix} 
\equiv \tilde{A}_h(\theta) \begin{pmatrix} u \\ v \end{pmatrix},
\end{equation}
where, putting everything together, we find
\begin{equation} \label{AtildeTheta}
\tilde{A}_h(\theta) =
\begin{pmatrix}
 \frac{-\sin{2\theta}}{\sqrt{4 h^2+4 h \cos{2 \theta}+1}}
 & \frac{(4 h+1) \left(\sqrt{4 h^2+4 h \cos{2\theta}+1}-2 h-\cos{2\theta}\right)-\sin^2{2\theta}}
 {\left(1-\cos{2\theta}\right)\sqrt{4 h^2+4 h \cos{2 \theta}+1}} \\ \vspace{0.1in}
 \frac{(4 h-1) \left(\sqrt{4 h^2+4 h \cos{2\theta}+1}+2 h+\cos{2\theta}\right)+\sin^2{2\theta}}
 {\left(1+\cos{2\theta}\right)\sqrt{4 h^2+4 h \cos{2 \theta}+1}}
 & \frac{\sin{2\theta}}{\sqrt{4 h^2+4 h \cos{2 \theta}+1}} \\
\end{pmatrix}.
\end{equation}
Thus, in order to understand the linear stability of leapfrogging orbits we must study this one-parameter family of non-autonomous two-by-two linear systems in which the coefficient matrices have period $T=\pi$.

\section{Review of Floquet theory}
\label{sec:floquet}
Floquet theory is concerned with exactly such problems, i.e., with systems of the form
\begin{equation} \label{floquet}
\dv{\z}{t} = B(t) \z, \qqtext{where} B(t+T) = B(t).
\end{equation}
Here $\z \in \Reals^n$ and $B(t)$ is an $n\times n$ matrix-valued function. The stability of the system is studied by considering its \emph{fundamental solution matrix} $\Phi(t)$, which sastisfies
$$
\dv{\Phi}{t} = B(t) \Phi, \, \Phi(0) = I,
$$
since, clearly, all solutions of Eq.~\eqref{floquet} are of the form $\Phi(t) \z_0$.
The \emph{monodromy matrix} is given by $M=\Phi(T)$. If $M$ has any eigenvalues $\lambda$ with $\abs{\lambda}>1$, then there exist solutions that grow exponentially in time. 

For Hamiltonian systems in dimension $n=2$, there is a useful diagnostic. For such systems $B(t) = J S(t)$ where $J = \left(\begin{smallmatrix} 0 & 1 \\ -1 &0 \end{smallmatrix} \right)$ and $S(t)$ is symmetric. Thus $\tr B(t)=0$, which implies that $\det \Phi(t)=1$ and, in particular, that $\det M=1$. The eigenvalues must then satisfy $\lambda_1\cdot\lambda_2=1$ and there are two generic cases:
\begin{enumerate}
\item If the eigenvalues are real, then, without loss of generality, we can choose $-1<\lambda_1 < 1$ and $\abs{\lambda_2}>1$, so that $\abs{\tr M}>2$, and the system is unstable
\item If the eigenvalues have nonzeros imaginary part then $\lambda_1 = \lambda_2^* = e^{i\theta}$, and $\tr M = 2 \cos{\theta}$ and, in particular $\abs{\tr M}<2$. The system is stable.
\end{enumerate}
In the borderline cases $\lambda_1 = \lambda_2 = \pm 1$, so that $\tr M = \pm 2$. In the case $\tr M = +2$, the system has a periodic solution with period $T$ and in the case $\tr M = -2$, the system has an anti-periodic solution with $\z(T) = -\z(0)$. The theory was developed by Floquet and is explained in more detail in many textbooks, for example that of Meiss~\cite{floquet1883equations,Meiss}.

\subsection{A perturbation expansion for the monodromy matrix}
\label{sec:monodromyPerturbation}
In what follows, we need to determine the stability of system of the form~\eqref{floquet} where
$$
B(t) = B_0(t) + \epsilon B_1(t),
$$
where we can assume for $\epsilon =0$, the system has fundamental solution matrix $\Phi_0(t)$ and monodromy matrix $M_0$. Letting
$$
\z = \Phi_0(t)\vb{w},
$$
then $\vb{w}$ solves
\begin{equation} \label{wdot}
\dv{\vb{w}}{t} = 
\epsilon \Phi_0^{-1}(t) B_1(t) \Phi_0(t) \vb{w} 
\equiv 
\epsilon \tilde{B}(t) \vb{w}.
\end{equation}
If system~\eqref{wdot} has fundamental solution matrix $\Phi_1(t)$, then system~\eqref{floquet} has fundamental solution matrix
$$
\Phi(t) = \Phi_1(t) \Phi_0(t)
$$
and monodromy matrix 
\begin{equation} \label{monodromyProduct}
M = \Phi_1(T)M_0.
\end{equation}

For the purposes of this paper, it suffices to calculate this term to leading order approximation in $\epsilon$, which we may compute as follows. Integrating in $t$, $\Phi_1$ solves
$$
\Phi_1(t) = I + \epsilon \int_{0}^{t} \tilde{B}(s) \Phi_1(s) \dd s.
$$
Picard iteration, a standard technique for showing the existence and uniqueness of solutions, see Ref.~\cite{Meiss}, can be used as an approximation method. Under additional assumptions, it generates a convergent sequence of approximations as the solutions to a recurrence relation
\begin{equation*}
\Psi_{n+1}  = I + \epsilon \int_{0}^{t} \tilde{B}(s) \Psi_n(s) \dd s, \qqtext{with}
\Psi_0  = I.
\end{equation*}
In order for this scheme to converge, the map being iterated must be a contraction on the space of continuous matrix-valued functions on $[0,T]$. Since $B(s)$ is continuous on the interval, it is bounded, and we may ensure it is a contraction  by choosing $\epsilon$ sufficiently small. The first iterate is
\begin{equation}\label{Psi1}
\Psi_1(t) = I + \epsilon \int_{0}^{t} \tilde{B}(s) \Psi_0(s) \dd s = I + \epsilon \int_{0}^{t} \tilde{B}(s) \dd s,
\end{equation}
which yields the approximation
\begin{equation} \label{Phi1}
 \Phi_1(t) = I + \epsilon \int_{0}^{t} \tilde{B}(s) \dd s + o(\epsilon).
\end{equation}

\section{Determining the stability}
\label{sec:analysis}
To prove the desired stability result, it suffices to show that the monodromy matrix of system~\eqref{uvlinearizedtheta} satisfies $\abs{\tr M}<2$ for $h<\eighth$ and $\abs{\tr M}>2$ for $h>\eighth$. In~\cite{BehringGoodman} we showed slightly less than this. First, we numerically computed a $\pi$-periodic solution to system~\eqref{uvlinearizedtheta} with $h=\eighth$, initially using MATLAB's \texttt{ode45} to within an error of $10^{-16}$, and then, to be doubly sure, using a method of order thirty and extended numerical precision in Julia, to an error of about $10^{-120}$. Second, we devised a perturbative procedure which approximates the value of $h$ at which a periodic orbit exists. This produced a sequence of approximations that converge exponentially to $h=\eighth$ in the order of the approximation.

Now, we complete the result. It turns out that for $h=\eighth$, system~\eqref{uvlinearizedtheta} has a closed form periodic orbit that can be expressed in terms of elementary functions. The coefficient matrix~\eqref{AtildeTheta} in this case is
\begin{equation} \label{AEighth}
\tilde{A}_\eighth =
\begin{pmatrix}
 -\frac{4 \sin {2\theta}}{\sqrt{8 \cos {2\theta}+17}} 
 & \frac{8 \cos^2{2\theta}-12 \cos {2\theta}+3 \sqrt{8 \cos {2\theta}+17}-11}{2 (1-\cos
   {2\theta}) \sqrt{8 \cos {2\theta}+17}} \\
 \frac{-8 \cos^2{2\theta}-4 \cos {2\theta}-\sqrt{8 \cos {2\theta}+17}+7}{2 (\cos {2\theta}+1) \sqrt{8 \cos {2\theta}+17}} 
 & \frac{4 \sin{2 \theta}}{\sqrt{8 \cos {2\theta}+17}} 
\end{pmatrix}.
\end{equation}
A periodic orbit $(u_1(\theta),v_1(\theta))$ with initial condition $(1,0)$ was found by entering the problem into the software package Maple, which returned the answer
\begin{equation} \label{uExact}
    \begin{pmatrix} u_1(\theta) \\ v_1(\theta) \end{pmatrix}
    = \frac{1}{20}
    \begin{pmatrix} 
        \phantom{\tan\theta (}
        1+4 \cos{2\theta}+3\sqrt{17+8 \cos {2\theta}}
         \\
        -\tan\theta \left(
        1+4 \cos{2\theta}+\sqrt{17+8 \cos{2\theta}}
        \right) 
        \end{pmatrix}. 
\end{equation}

For the next part of the argument, we need to find the fundamental solution matrix $\Phi$ for system~\eqref{uvlinearizedtheta}. The solution~\eqref{uExact} forms the first column of $\Phi$. We may find the second column of $\Phi$, i.e., a solution $(u_2(\theta),v_2(\theta))$ with initial condition $(0,1)$ by reduction of order. Abel's identity ensures that, because $\tr \tilde{A}_\eighth =0$, the fundamental solution matrix satisfies $\det\Phi = 1$, which we use to find
$$
v_2(\theta) = \frac{1+v_1(\theta) u_2(\theta)}{u_1(\theta)}.
$$
Substituting this into system~\eqref{uvlinearizedtheta} gives a nonhomogeneous first-order equation for $u_2(\theta)$, which we integrate to find
$$
\begin{pmatrix} u_2(\theta) \\ v_2(\theta) \end{pmatrix} = 
\begin{pmatrix} u_2^{\rm p} (\theta) \\ v_2^{\rm p} (\theta) \end{pmatrix} + 
\begin{pmatrix} u_2^{\rm np} (\theta) \\ v_2^{\rm np} (\theta) \end{pmatrix},
$$
where the periodic part is given by
$$
\begin{pmatrix} u_2^{\rm p} (\theta) \\ v_2^{\rm p} (\theta) \end{pmatrix} =
\begin{pmatrix}
\frac{-3 (1672 \cos{\theta}+1801 \cos{3 \theta}+321 \cos{5 \theta}-44 \cos{7 \theta })+ (2794 \cos{\theta}-323 \cos{3 \theta}-243 \cos{5 \theta}+22 \cos{7 \theta})\sqrt{8 \cos{2 \theta}+17}}{150 (-28 \sin{\theta}-9 \sin{3 \theta}+\sin{5 \theta })} \\
\frac{1}{150} \left( 13 -88 \cos{2 \theta}+(89-44 \cos{2 \theta}\right) \sqrt{8 \cos{2 \theta}+17})
\end{pmatrix},
$$
and the nonperiodic part is given by
$$
 \begin{pmatrix} u_2^{\rm np} (\theta) \\ v_2^{\rm np} (\theta) \end{pmatrix} =
 \left(
   \tfrac{22}{3} E\left(\theta\left|\tfrac{16}{25}\right.\right) 
   - 2 F\left(\theta\left|\tfrac{16}{25}\right.\right)
  \right) 
 \begin{pmatrix} u_1(\theta) \\ v_1(\theta) \end{pmatrix}.
$$
Here, $E(\theta \lvert m)$ and $F(\theta \lvert m)$ are incomplete elliptic integrals with parameter $m$ of the second and first kind, respectively~\cite[\S19]{NIST:DLMF}. Each of these grows, on average, linearly in $\theta$, demonstrating that this solution is not periodic. Evaluating $\Phi$ at $\theta=\pi$ yields the monodromy matrix
\begin{equation}\label{M0}
M= \begin{pmatrix} 
1 & \mu\\
0 & 1
\end{pmatrix},
\end{equation}
where 
$$
\mu =\frac{44}{3} E\left(\frac{16}{25}\right)-4 K\left(\frac{16}{25}\right) \approx 10.74,
$$
and $E(m)=E\left(\left.\frac{\pi}{2}\right\rvert m\right) $ and $K(m)=F\left(\left.\frac{\pi}{2}\right\rvert m\right)$ are here \emph{complete} elliptic integrals of the second and first kind, respectively.

Now we consider the case that $h = \eighth + \epsilon$, and expand the coefficient matrix $\tilde{A}_h$ in powers of $\epsilon$,
$$
\tilde{A}_{\eighth+\epsilon} = 
A_0(\theta) + \epsilon A_1(\theta) + o(\epsilon),
$$
where $A_0 = \tilde{A}_{\eighth}$ is given by Eq.~\eqref{AEighth}
and
$$
A_1 = 
\begin{pmatrix}
 \frac{32 (\sin{2 \theta}+2 \sin{4 \theta})}{(8 \cos{2 \theta}+17)^{3/2}} 
 & 2 \csc^2{\theta}\left(\frac{ -68 \cos{2 \theta}+4 \cos{4 \theta}-8 \cos{6 \theta}-53}{(8 \cos{2 \theta}+17)^{3/2}}+1\right)  \\
 2 \sec^2{\theta} \left(\frac{ 68 \cos{2 \theta}+28 \cos{4 \theta}+8 \cos{6 \theta}+21}{(8 \cos{2 \theta}+17)^{3/2}}+1\right)  & 
 -\frac{32 (\sin{2 \theta}+2 \sin{4 \theta})}{(8 \cos{2 \theta}+17)^{3/2}} \\
\end{pmatrix}
$$
and apply the perturbation argument of Sec.~\ref{sec:monodromyPerturbation}.

Calculating the trace of the monodromy matrix~\eqref{monodromyProduct} with leading-order term given by Eq.~\eqref{M0} shows that 
\begin{equation}\label{traceOfM}
\tr M = 2 + \mu \Phi_1(\pi)_{2,1},
\end{equation}
and our remaining task is to find the $(2,1)$ entry of $\Phi_1(\pi)$. In particular, the needed element of $\Phi_1$ in Eq.~\eqref{Phi1}
$$
\Phi_1(\pi)_{2,1} = I + \epsilon \int_0^{\pi} \tilde{B}_{2,1}(\theta) \dd \theta,
$$
where
\begin{multline*}
\tilde{B}_{2,1}(\theta) = 
   \frac{2 \sec^2{\theta}}{25}
   \Bigg(
    \frac{256 \cos{2 \theta}+100 \cos{4 \theta}+16 \cos{6 \theta}+253}{8 \cos{2 \theta}+17} \\
    + \frac{1536 \cos{2 \theta}+660 \cos{4 \theta}+148 \cos{6 \theta}+12 \cos{8 \theta}+769}{(8 \cos{2 \theta}+17)^{3/2}}
   \Bigg).
\end{multline*}
Integrating gives
$$
\Psi_1(\pi)_{2,1} =
\frac{4}{5} \left(11 E\left(\tfrac{16}{25}\right)-3 K\left(\tfrac{16}{25}\right)\right) \epsilon \approx 6.44 \epsilon .
$$

In particular, we have found that both $\mu>0$ and $\Psi_1(\pi)_{2,1}>0$. Therefore, by Eq.~\eqref{traceOfM}, $\tr M >0$ when $\epsilon>0$ and the leapfrogging orbit is unstable, and $\tr M<0$ when $\epsilon<0$ and the leapfrogging orbit is stable. Thus we have resolved Tophøj and Aref's conjecture from Ref.~\cite{Tophoj2013} that a bifurcation occurs at $\alpha = \phi^{-2}$. 

That all of this work still seems miraculous to these authors, and the mystery remains as to why.

\section*{Acknowledgments}
We dedicate this article to the memory of Denis Blackmore, an esteemed colleague who served with wit and generosity on the dissertation committee of BMB. This research arises from the dissertation research of BMB, supported by NJIT. RG gratefully acknowledges support from the NSF under DMS–2206016. We thank Panos Kevrekidis, who insisted that the question of linearized stability was not fully answered in~\cite{BehringGoodman} and encouraged us to do the perturbation theory properly. Finally, we must acknowledge how the solution~\eqref{uExact} was found. We tried and failed to to find an exact the initial value problem~\eqref{uvlinearized} with $h=\frac{1}{8}$ using both Mathematica and pencil and paper. We subsequently posted a question to the Math Overflow discussion board to ask for help. Steve Israel responded with a solution to the equation he found using Maple~\cite{mathoverflow}. We are very grateful to him, and we still wonder how Maple solved it. All other difficult integrals performed in this note were accomplished using Mathematica, without which this work would have been impossible.

\bibliography{leapfrog}
\bibliographystyle{elsarticle-num} 

\end{document}